# A Silicon-Based Monolithic Optical Frequency Comb Source

Mark A. Foster[1], Jacob S. Levy[2], Onur Kuzucu[1], Kasturi Saha[1], Michal Lipson[2], & Alexander L. Gaeta[1]

[1]*School of Applied and Engineering Physics, Cornell University, Ithaca, NY 14853, USA*

[2]*School of Electrical and Computer Engineering, Cornell University, Ithaca, NY 14853, USA*

**Recently developed techniques for generating precisely equidistant optical frequencies over broad wavelength ranges are revolutionizing precision physical measurement [1-3]. These frequency "combs" are produced primarily using relatively large, ultrafast laser systems. However, recent research has shown that broad-bandwidth combs can be produced using highly-nonlinear interactions in microresonator optical parametric oscillators [4-11]. Such devices not only offer the potential for developing extremely compact optical atomic clocks but are also promising for astronomical spectroscopy [12-14], ultrashort pulse shaping [15], and ultrahigh-speed communications systems. Here we demonstrate the generation of broad-bandwidth optical frequency combs from a CMOS-compatible integrated microresonator [16,17], which is a fully-monolithic and sealed chip-scale device making it insensitive to the surrounding environment. We characterize the comb quality using a novel self-referencing method and verify that the comb line frequencies are equidistant over a bandwidth that is nearly an order of magnitude larger than previous measurements. In addition, we investigate the ultrafast**

**temporal properties of the comb and demonstrate its potential to serve as a chip-scale source of ultrafast (sub-ps) pulses.**

Optical frequency combs enable the precise measurement of optical frequencies through direct referencing to microwave atomic clocks. Future atomic clockwork is expected to utilize optical frequency combs to transfer the rapid oscillations of optical frequency standards to measurable microwave frequency ranges [1-3]. Combs are traditionally generated using mode-locked ultrafast laser sources. Recently, the generation of optical frequency combs through the nonlinear process of continuous-wave optical-parametric oscillation using micro-scale resonators has attracted significant interest [4-10] since these devices have the potential to yield highly compact and frequency agile comb sources. The microresonator structures include silica microtoroids, silica microspheres, silica-fibre Fabry-Perot cavities, and $CaF_2$ microresonators [4-10]. These demonstrations have yielded impressive results including the generation of spectral bandwidths up to 150 THz (octave-spanning) with a comb spacing of 850 GHz [10], as well as with spacings as small as 14 GHz but with a reduced bandwidth of 3.7 THz [7]. The frequency spacing of the comb generated in microtoroids was found to be equidistant to $7 \times 10^{-18}$ relative to the optical frequency [4]. Additionally, stabilization of a microtoroid comb to a microwave source was demonstrated by controlling the pump laser power and frequency in a feedback loop [5].

While the full dynamics of parametric comb generation is highly complex and resonator specific [9,18-20], the fundamental mechanism is shared by all geometries. This mechanism relies on a combination of optical parametric amplification and oscillation as a result of the nonlinear optical process of four-wave mixing (FWM) within the resonator [21-23]. Specifically, a single pump laser is tuned to a cavity resonance, and the resulting parametric amplification provides optical gain for the surrounding spectral modes. With sufficient pump power the amplification exceeds the round-trip loss and some of the modes undergo oscillation. The presence of multiple oscillating frequencies leads to FWM that generates light in additional modes. This frequency conversion combined with parametric amplification can lead to a vast cascading of the oscillating frequencies and due to energy conservation the newly generated frequencies are precisely equidistant.

Here we demonstrate the generation of optical frequency combs from a highly-robust CMOS-compatible integrated microresonator optical parametric oscillator [16,17]. Both the microresonator and the coupling waveguide are fabricated monolithically in a single silicon nitride layer using electron-beam lithography and subsequently clad with silica. Existing microresonator comb sources require coupling with fragile and mechanically sensitive tapered or cleaved optical fibres and operation in sealed environments to avoid contaminants, whereas our system yields a fully-monolithic and sealed device with coupling and operation that is insensitive to the surrounding environment.

An example of the type of structures we employ is the 112-μm-radius microring resonator shown in Fig. 1(a). The free-spectral range of the resonator is 204 GHz, and the loaded quality factor $Q$ of the resonator is measured to be $3 \times 10^5$. The cross-sectional dimensions of the coupling waveguide and the ring are both given by 750-nm tall by 1500-nm wide. These dimensions are carefully designed to produce anomalous group-velocity dispersion at the pump wavelength of 1550 nm [23,24], and the microring is fabricated as previously described [16]. We inject 300 mW from a tuneable diode laser amplified by an erbium-doped fibre amplifier into the coupling waveguide. The polarization of this pump laser is set to the quasi-TE mode of the waveguide, and the laser is tuned onto resonance such that a stable "thermal-lock" is achieved [5]. As the pump laser is tuned into resonance, the parametric oscillation threshold is reached and cascaded FWM generates an array of new frequencies. Further increase of the pump power within the resonator yields a greater spectral extent and density of comb, and an example spectrum is shown in Fig 1(b). We observe the generation of more than 350 comb lines spanning over 75 THz (1375 nm to 2100 nm) and spaced by 204 GHz.

For applications in metrology, the generated optical frequencies must be precisely equidistant. Lines generated through cascaded FWM obey this spacing. However, components can also arise from independent oscillations and may not lie on this equidistant comb. Therefore, it is critical to verify that the generated lines represent an equidistant optical frequency comb. To test for this equidistance, we first implement multi-heterodyne beat-note detection with a 38-MHz reference comb from a mode-locked laser. This technique was previously used to characterize the equidistance of silica microtoroid combs [4]. The specifics of our implementation are described in the methods section. Essentially, six lines of the microresonator comb are combined with a reference comb on a photodetector. The lines from the two combs generate a series of beat notes, which we measure using an RF spectrum analyser. The equidistance of the observed RF tones verifies that the comb lines are uniformly spaced in frequency. The generated RF-spectrum for our measurement is shown in Fig. 2(b), and we find the beat notes equidistant to within the measurement accuracy of 10 kHz. This corresponds to better than $9.7 \times 10^{-9}$ relative to the spectral range of the measurement and better than $5.2 \times 10^{-11}$ relative to the optical frequency. The results of this measurement are comparable to those carried out on silica-microtoroids [4].

To increase the precision of the measurements, we replace the spectrum analyser by a frequency counter, as was previously implemented with silica-microtoroids by comparing 3 comb lines individually with a locked reference comb [4]. Here we develop a new self-referencing technique, which we term parametric comb folding. This method alleviates the need for a locked reference comb and facilitates the characterization of a large number of lines (70 lines here) over a substantially greater bandwidth (> 100 nm) to a level of precision comparable to previous techniques.  The details of parametric comb folding are described in the methods section. In essence, the method uses a continuous-wave transfer laser and parametric frequency conversion in a highly nonlinear fibre to test the symmetry of the comb about the transfer laser frequency. The transfer laser is offset-locked to one of the lines of the comb under test, and the parametric interaction with the comb under test effectively folds the comb in half about the transfer laser frequency. If the comb under test is symmetric about this frequency, then the beat notes measured between the original lines and the folded lines will be twice that between the transfer laser and the line to which it is locked. Comparing these frequencies using a counter in ratio mode allows for the precise measurement of the comb symmetry. To determine that the comb is not simply symmetric but is equidistant, the symmetry is tested about two more positions (7 and 10 comb lines away here). Three measurement points with no common factors in the separation is critical to eliminate the possibility of periodic symmetry. The results of this measurement are shown in Fig. 3. The deviation from symmetry is shown for three different measurement frequencies. The centre of the comb (1545 nm) and 7 and 10 comb lines away (1557 nm and 1562 nm). In all cases, the deviation from symmetry is less than 0.5 Hz over the 115-nm (14.5 THz) span of the measurement. This corresponds to a comb equidistance of $3\times10^{-14}$ relative to the measurement range and $3\times10^{-15}$ relative to the optical frequency.

For precision measurement, it is desirable to stabilize the DC-offset frequency of the optical frequency comb. The most common method by which to stabilize this parameter is to implement a self-referencing scheme in the form of an *f-2f* interferometer. This method requires the frequency comb to span an octave in bandwidth, an extent which was recently demonstrated in silica microtoroids [10]. The current 75-THz span is sufficient to implement a *2f-3f* interferometer [26], and we expect to achieve octave-spanning bandwidth in future CMOS-compatible devices by a combination of improved dispersion engineering [25] and increased pump power.

As a final demonstration, we investigate the temporal properties of the generated comb. A mode-locked laser source produces a periodic train of ultrafast pulses at a rate given by the comb spacing. We expect that if the comb lines of our microresonator have a definite phase relationship, the its output should consist of a periodic ultrafast waveform in the time domain. To investigate the temporal nature of our generated comb, we filter out a 10-nm bandwidth at 1560-nm and send it to a fibre amplifier and auto-correlator. We find that each time the pump laser is tuned onto resonance a different temporal output is generated. However, the temporal waveform is stable as long as the pump is maintained on resonance. Figure 4(a) shows the autocorrelation of the comb section for various realizations. Intriguingly, in some cases the generation of isolated ultrafast pulses is observed. For example, as seen in Fig. 4(b), we observe the generation of 740-fs pulses with a 204-GHz repetition rate. We anticipate that this pulsed mode can be favoured and therefore robustly achieved with the incorporation of a mechanism for saturable absorption. Furthermore, using the full comb bandwidth will potentially allow for the generation of pulses approaching 10 fs.

In summary, we have demonstrated the generation of broad-bandwidth (75-THz) optical frequency combs from a highly-robust and compact CMOS-compatible microresonator. We developed a new technique to precisely characterize the generated comb and verified the mode equidistance of a 115-nm (14.5 THz) span to better than $3 \times 10^{-15}$ relative to the optical frequency. Additionally, we investigate the ultrafast temporal properties of a filtered portion of the microresonator comb and observe the generation of 740-fs pulses at a repetition rate of 204 GHz. Based on these results and the compatibility with CMOS electronics, this extremely compact source can serve as a fully chip-scale replacement for relatively bulky mode-locked lasers in many precision frequency metrology and ultrafast photonic applications. In particular, this robust source is ideal for applications where high-repetition rates are desired such as the calibration of astronomical spectrographs [12-14], line-by-line pulse shaping [15], and ultra-high-speed communications systems.

**Methods**

We implement multi-heterodyne beat note detection as follows. The pump laser is tuned onto resonance such that a dense optical frequency comb is generated, and the pump laser is then locked to a single comb line of the 38-MHz reference comb. We filter out six consecutive generated lines from the microresonator comb in the range of 1555 nm to 1564 nm as shown in Fig. 2(a). These six lines are combined with the reference comb and the beat notes are detected using a 10-GHz-bandwidth photodetector. If the spacing of the microresonator comb is nearly but not exactly an integer multiple of the reference comb spacing, then the RF spectrum measured from the photodetector will show a series of six beat-notes. These tones correspond to the difference in frequency between each of the microresonator comb lines and a nearby tooth of the reference comb. Therefore, the equidistance of the observed RF tones verifies that the comb lines are uniformly spaced in frequency.

Parametric comb folding is implemented as follows. A continuous-wave transfer laser is offset-locked to one of the lines of the microresonator comb under test. This transfer laser is amplified and combined with the comb under test in a highly nonlinear fibre. The strong transfer laser serves as a degenerate pump for four-wave-mixing frequency conversion in the nonlinear fibre. This process converts all the comb lines from one side of the pump to the other side (and vice-versa) such that the newly generated lines are symmetric about the pump (see Fig. 3a). In this way, the transfer laser effectively folds the comb in half about the transfer laser frequency. If the comb under test is symmetric about this frequency then the beat notes measured between the original lines and the generated lines will be twice the beat note between the transfer laser and the comb line it is locked with. Comparing these frequencies using a frequency counter in ratio mode allows for the precise measurement of the comb symmetry about this frequency. To determine that the comb not simply symmetric but is in fact equidistant, we measure the symmetry about two more frequencies (7 and 10 comb lines away from this initial frequency). Three measurement points with no common factors in the separation are required to eliminate the possibility of periodic symmetry.

This work was supported by DARPA under the POPS and ZOE programs and by the Center for Nanoscale Systems, supported by the NSF and the New York State Office of Science, Technology and Academic Research. This work was performed in part at the Cornell NanoScale Facility, a member of the National Nanotechnology Infrastructure Network, which is supported by the National Science Foundation (Grant ECS-0335765).

The authors declare no competing financial interests. Correspondence and requests for materials should be addressed to A.L.G. (a.gaeta@cornell.edu).

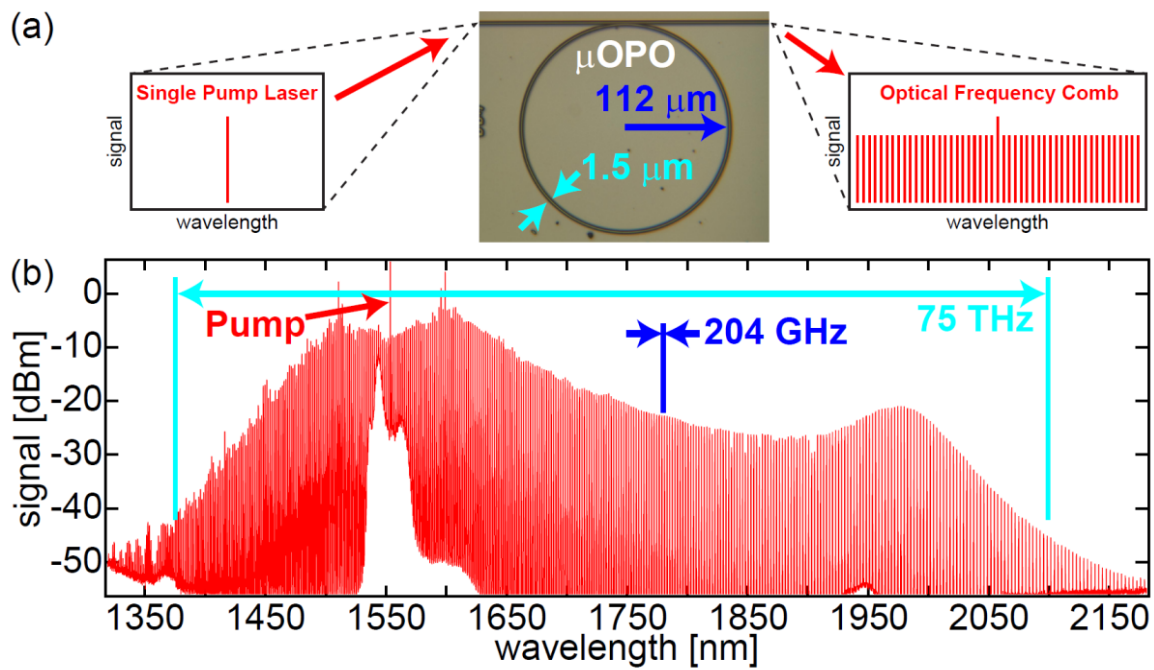

**Figure 1 | CMOS-Compatible Optical Frequency Comb Source. a,** A single pump laser is coupled into the CMOS-compatible silicon nitride microresonator (μOPO) shown in the optical micrograph. A highly nonlinear interaction resulting in four-wave mixing and optical parametric oscillation leads to a vast multiplication of the oscillating frequencies and the generation of an optical frequency comb. **b,** Experimentally measured optical frequency comb generated in this device. The comb spans 75 THz and has a line spacing of 204 GHz, which corresponds to more than 350 lines that are generated.

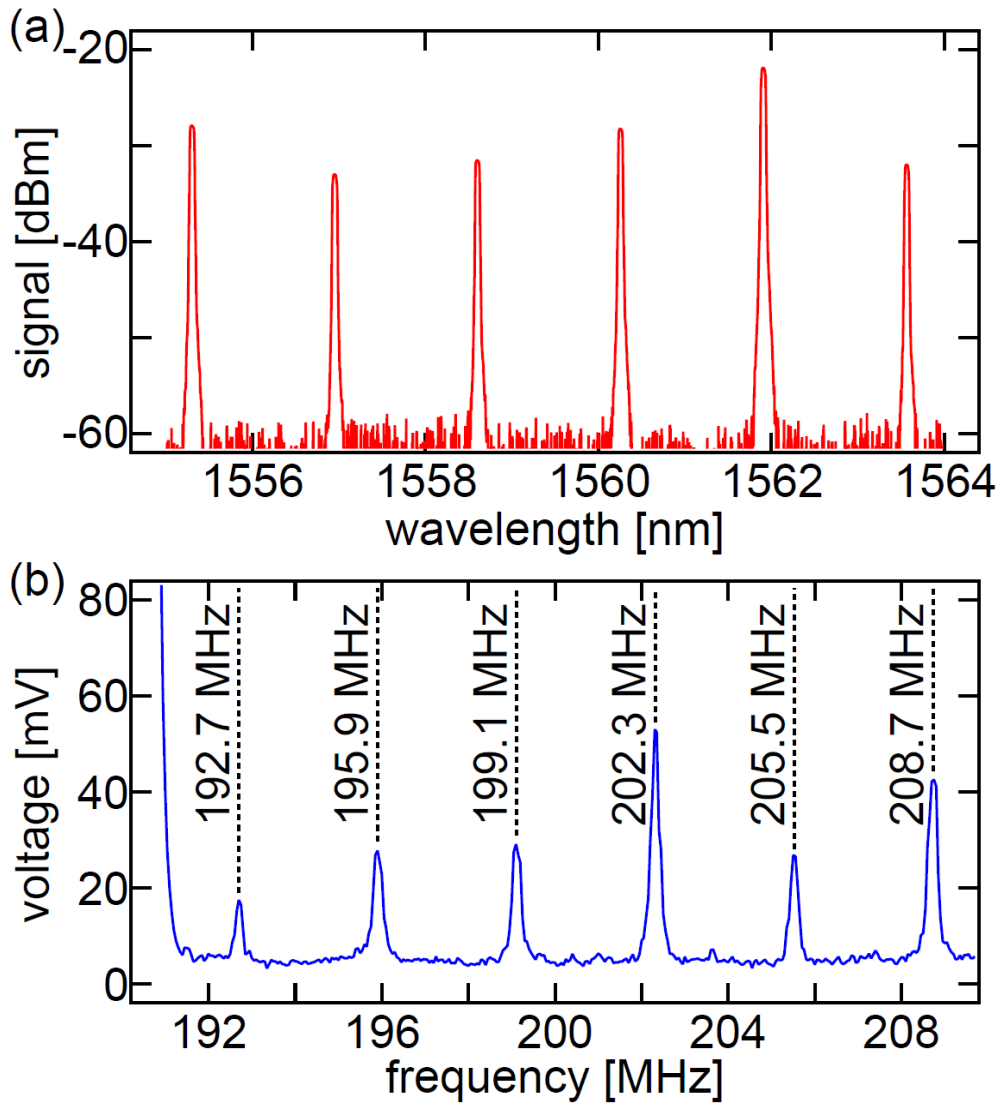

**Figure 2 | Multi-Heterodyne Beat Note Detection. a,** Six consecutive comb lines from 1555 nm to 1564 nm are used to perform multi-heterodyne beat-note detection with a 38-MHz reference comb from a mode-locked fibre laser. **b,** The six heterodyne beat notes are measured simultaneously with an RF spectrum analyser. The equidistance of the RF beat notes verifies the equidistance of the six comb modes to within the measurement accuracy of 10 kHz.

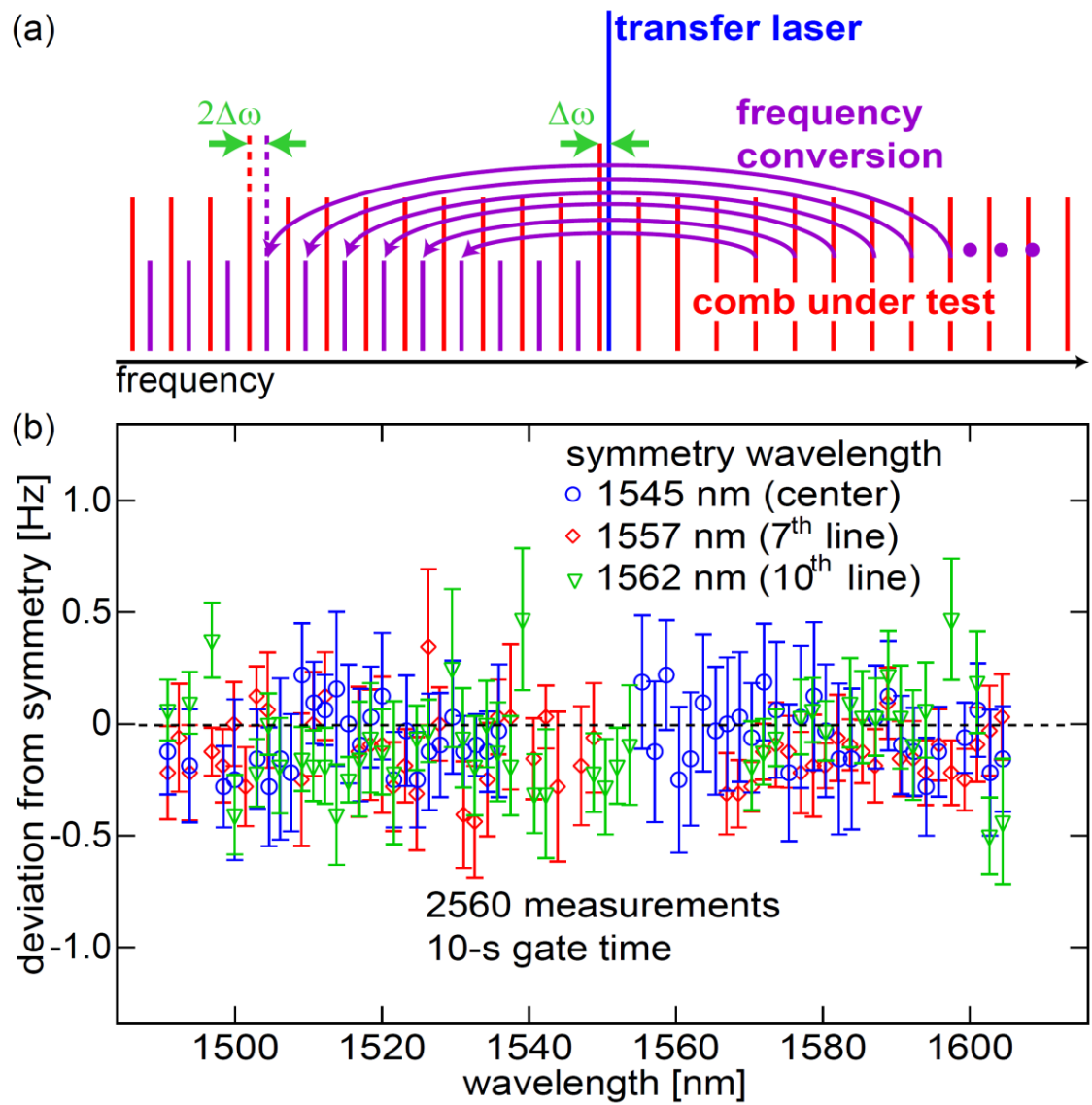

**Figure 3 | Parametric Comb Folding. a,** A strong transfer laser is locked to one line of the comb under test. The comb and the transfer laser are combined and sent through a highly nonlinear fibre where four-wave mixing frequency conversion takes place. This frequency conversion process converts the lines on one side of the transfer laser to the other and vice versa, thereby effectively folding the comb about the transfer laser frequency. By comparing the beat frequency between the original comb lines and the folded lines, the symmetry of the comb about the transfer frequency is determined with high precision. **b,** Experimental symmetry measurements for the transfer laser locked to the comb centre, 7th line from centre and 10th line from centre. In all cases the comb is symmetric to better than 0.5 Hz over the full 115-nm range of the measurement. These three measurement points with no common factors eliminate the possibility of simple symmetry and periodic symmetry indicating that the lines are indeed equidistant. This characterization involved 2560 individual measurements on 70 comb lines.

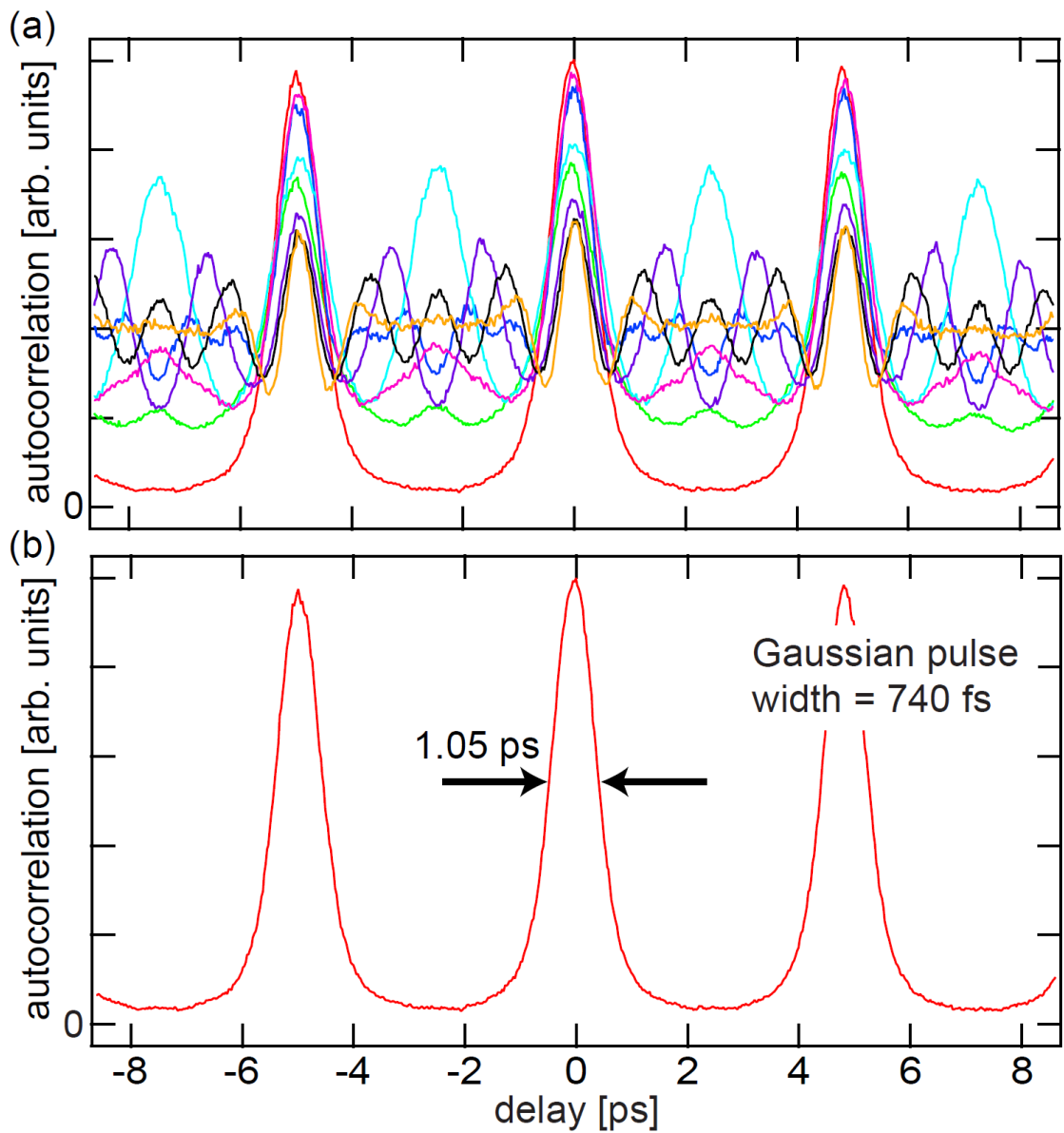

**Figure 4 | Ultrafast Temporal Characterization of Generated Comb.** A 10-nm bandwidth of the generated comb is characterized with an ultrafast autocorrelator. **a,** The autocorrelation of the comb section for various initialisations of the OPO. We find that each time the pump laser is tuned onto resonance with the microresonator a different periodic ultrafast temporal waveform is generated. This waveform is stable as long as the pump is maintained on resonance. **b,** In some cases, isolated ultrafast pulses are generated. We observe the generation of 740-fs pulses with a 204-GHz repetition rate assuming a Gaussian pulse shape.